# WIENER RECONSTRUCTION OF GALAXY SURVEYS IN SPHERICAL HARMONICS

O. Lahav[1], K.B. Fisher[1], Y. Hoffman[2], C. A. Scharf[1] & S. Zaroubi[2]

(1) Institute of Astronomy, Madingley Rd., Cambridge CB3 0HA, UK

(2) Racah Institute of Physics, The Hebrew University, Jerusalem, Israel

**ABSTRACT.** The analysis of whole-sky galaxy surveys commonly suffers from the problems of shot-noise and incomplete sky coverage (e.g. at the Zone of Avoidance). The orthogonal set of spherical harmonics is utilized here to expand the observed galaxy distribution. We show that in the framework of Bayesian statistics and Gaussian random fields the $4\pi$ harmonics can be recovered and the shot-noise can be removed, giving the most probable picture of the underlying density field. The correction factor from observed to reconstructed harmonics turns out to be the well-known Wiener filter (the ratio of signal to signal+noise), which is also derived by requiring minimum variance. We apply the method to the projected 1.2 Jy IRAS survey. The reconstruction confirms the connectivity of the Supergalactic Plane across the Galactic Plane (at Galactic longitude $l \sim 135^o$ and $l \sim 315^o$) and the Puppis cluster behind the Galactic Plane ($l \sim 240^o$). The method can be extended to 3-D in both real and redshift space, and applied to other cosmic phenomena such as the COBE Microwave Background maps.

*Subject Headings: galaxies - large scale structure of the universe - surveys - data analysis*

## 1. INTRODUCTION

Two basic problems commonly appear in analysing the distribution of galaxies. First, if one assumes that the distribution of luminous galaxies samples an underlying smooth density field, then the discreteness of objects introduces Poisson 'shot-noise'. Second, incomplete sky coverage, e.g. due to the obscuration by the Galactic Plane (the Zone of Avoidance), is an obstacle in mapping the whole-sky distribution. In this *Letter* we show how to recover the all-sky projected density field, characterized by an assumed power-spectrum of fluctuations, from a galaxy survey which suffers incomplete sky described by a known mask.

The recovery of a signal from noisy and incomplete data is a classic problem of inversion. A straightforward inversion is often unstable, and a regularization scheme of some sort is essential in order to interpolate where data are missing or noisy. In the Bayesian spirit we use here raw data and a prior model to produce 'improved data'. The prior model does not necessarily require a speculative assumption. In the context of this work we simply require a reconstruction which obeys the constraint of the 2-point correlation function of the observed galaxy distribution, as derived from a smaller section of the sky. Using the above principles we derive a Wiener filter (the ratio of signal to signal+noise), which also follows from requiring minimum variance (e.g. Rybicki & Press 1992). There are many (related) variants of this approach, including Maximum Entropy (e.g. Gull 1989). The reconstruction problems can be addressed within the framework of conditional probability and constrained realizations of Gaussian random fields (Bertschinger 1987, Binney & Quinn 1990, Hoffman & Ribak 1991). This has been applied to reconstruct the density field from the observed peculiar velocities (Kaiser & Stebbins 1991, Stebbins 1993, Ganon & Hoffman 1993).

A simple example of a Wiener filter is given in Section 2, and the formulation for the whole-sky reconstruction in spherical harmonics is shown in Section 3. Section 4 shows application to the projected 1.2 Jy IRAS survey, and future work is discussed in Section 5. We shall give the full mathematical details and extension of the method to 3-D harmonics elsewhere (Zaroubi et al., in preparation).

## 2. A SIMPLE EXAMPLE OF WIENER FILTER

Let us first consider a simple pedagogical example. Assume two Gaussian variables, $x$ and $y$, with zero mean, $\langle x \rangle = \langle y \rangle = 0$ (hereafter $\langle ... \rangle$ denote ensemble average). The probability for $x$ *given* $y$ is by the rule of conditional probability

$$P(x|y) = \frac{P(x,y)}{P(y)}, \qquad (1)$$

where $P(y)$ is a 1-dimensional Gaussian probability distribution, and the joint probability $P(x,y)$ is a bi-variate Gaussian distribution. It is straightforward to show that $P(x|y)$ is a 'shifted Gaussian' with the maximum probability occurring for $\hat{x} = \frac{\langle xy \rangle}{\langle y^2 \rangle} y$. In the special case of Gaussian fields the most probable reconstruction is also the mean field. Hereafter we term them together as the 'optimal reconstruction'.

Exactly the same result for the 'optimal reconstruction' is also obtained by a different approach, by asking for the linear filter $F$ which minimizes the variance $\langle (x - Fy)^2 \rangle$. Minimizing with respect to $F$ gives indeed $\hat{F} = \frac{\langle xy \rangle}{\langle y^2 \rangle}$ and $\hat{x} = \hat{F}y$, as above. Note that although the results of the two approaches are identical, due to the quadratic nature and linearity of filter of the functions involved, the assumptions involved are quite different. The conditional probability approach (eq. 2) requires to specify the full distribution functions (Gaussians in our case). On the other hand, the minimum variance approach only considers the variance (second moment) of the distribution function, but assumes a linear filter $F$.

Consider now the special case that $y = x + \sigma$, where $\sigma$ is a Gaussian noise uncorrelated with the true signal $x$ (hence $\langle x\sigma \rangle = 0$). It follows that the optimal estimator of the signal $\hat{x}$ given the (noisy) measurement $y$ is

$$\hat{x} = \frac{\langle x^2 \rangle}{\langle x^2 \rangle + \langle \sigma^2 \rangle} y \qquad (2)$$

The factor $(F)$ in front of the measurement $y$ is the well-known Wiener filter commonly used in signal processing (for review see e.g. Press et al. 1992, Rybicki & Press 1992). Note that it requires *a priori* knowledge of the variances in the signal and the noise. When the noise is negligible the factor approaches unity, but when it is significant the measurement is attenuated.

A third approach to the problem, in which the Bayesian *a posteriori* probability (cf. eq. 2), $-\ln P(x|y) = -[\ln P(y|x) + \ln P(x)] = (y-x)^2/\langle\sigma^2\rangle + x^2/\langle x^2\rangle$, is maximized with respect to $x$, also yields the same result as the other two approaches. In fact, this is a special case of the $\chi^2$ (log-Likelihood) minimization subject to regularization of the form $\chi^2 + \alpha f(x)$, where f(x) is the regularizing function (e.g. $f(x) = x^2$ in our case, and in other applications taken as the 'entropy' $f(x) = x \ln x$), and $\alpha$ is a Lagrange multiplier. We see that a regularization with a prior $f(x) = x^2$ is essentially equivalent to a Wiener filter.

## 3. NOISE REMOVAL AND MASK INVERSION

For simplicity, we shall consider here projected (2-dimensional) galaxy samples. We formulate our problem as follows: What are the full-sky noise-free harmonics given the observed harmonics, the mask describing the unobserved region, and a prior model for the power-spectrum of fluctuations ?

### 3.1. Expansion in Spherical Harmonics

Here we use spherical harmonics to expand the galaxy distribution in a whole-sky survey. This technique has been considered for 2-D samples (e.g. Peebles 1973, Scharf et al. 1992) and more recently for analysing redshift surveys (Scharf & Lahav 1993; Lahav et al. 1993; Fisher, Scharf & Lahav 1993b). Basically, the projected density field over $4\pi$ is expanded as a sum:

$$S(\theta, \phi) = \sum_{l} \sum_{m=-l}^{m=+l} a_{lm} Y_{lm}(\theta, \phi), \qquad (3)$$

where the $Y_{lm}$'s are the orthonormal set of spherical harmonics. A reconstruction up to harmonic $l_{max}$ resolves structure on angular scale of $\pi/l_{max}$. The spherical harmonic analysis provides a unified language to describe the local cosmography as well as the statistical properties (e.g. power-spectrum) of the galaxy distribution.

### 3.2. Mask Inversion Using Wiener Filter

We turn now to the more complicated problem of the harmonics with incomplete sky coverage. Here we consider a 'sharp' mask, in which observed regions are assigned equal weight, while masked regions are assigned zero weight. The observed harmonics $c_{lm,obs}$ (with the masked regions filled in uniformly according to the mean) are related to the underlying 'true' whole-sky harmonics $a_{lm}$ by (cf. Peebles 1980, eq. 46.33)

$$c_{lm,obs} = \sum_{l'} \sum_{m'} W_{ll'}^{mm'} \left[ a_{l'm'} + \sigma_a \right] \tag{4}$$

where the monopole term ($l' = 0$) is excluded. We have added the shot-noise $\sigma_a$ in the 'true' number-weighted harmonics $a_{lm}$'s (not in the $c_{lm}$'s). The noise variance is estimated as $\langle \sigma_a^2 \rangle = \mathcal{N}$ (the mean number of galaxies per steradian, independent $l$ in this case). The harmonic transform of the mask, $W_{ll'}^{mm'}$, introduces 'cross-talk' between the different harmonics and acts as a 'point spread function' (in analogy with problems in image processing).

By the rule of conditional probability (cf. eq. 1)

$$P(\mathbf{a}|\mathbf{c}_{obs}) = \frac{P_G(\mathbf{a}, \mathbf{c}_{obs})}{P_G(\mathbf{c}_{obs})} , \tag{5}$$

where the vectors $\mathbf{a}$ and $\mathbf{c}_{obs}$ represent the sets of observed harmonics $\{a_{lm}\}$ and $\{c_{lm,obs}\}$, and $P_G$ stands for an assumed Gaussian distribution function with variance and covariance which depend on an assumed power-spectrum. As in the simple example above one can now find the estimator $\hat{\mathbf{a}}$ which gives maximum probability. This is a special case of a constrained realizations formalism (Hoffman & Ribak 1991), but here the formulation and computation are greatly simplified due to the orthogonality of the harmonics. An alternative approach is finding the minimum of the variance $\langle |\mathbf{a} - \mathbf{F}\mathbf{W}^{-1}\mathbf{c}_{obs}|^2 \rangle$ with respect to a desired Wiener filter matrix $\mathbf{F}$ (here $\mathbf{W}^{-1}$ stands for the 'pseudo inverse' of $\mathbf{W} \equiv \{W_{ll'}^{mm'}\}$). We shall present the full derivation by the two approaches elsewhere (Zaroubi et al., in preparation), but the answer for the optimal reconstruction can be seen by analogy with the simple example given above (eq. 2) :

$$\hat{\mathbf{a}} = \mathbf{F}\mathbf{W}^{-1}\,\mathbf{c}_{obs}, \qquad (6)$$

with the diagonal matrix

$$\mathbf{F} = diag\left\{\frac{\langle a_l^2 \rangle_{th}}{\langle a_l^2 \rangle_{th} + \langle \sigma_a^2 \rangle}\right\}. \qquad (7)$$

Only the diagonal elements appear in the $\mathbf{F}$ matrix due to the orthogonality of the harmonics. We emphasize again that in the case of underlying Gaussian field the most probable field, the mean field and the minimum variance Wiener filter are all identical. Hence these different approaches are unified.

Even if the sky coverage is $4\pi$ ($\mathbf{W} = \mathbf{I}$), the Wiener filter is essential to reveal the optimal underlying 'continuous' density field, cleaned of noise. The correction factor is per $l$, independent of $m$, so in the case of full sky coverage, only the amplitudes are affected by the correction, but not the relative phases. For example, the dipole direction is not affected by the shot-noise, only its amplitude. But of course, if the sky coverage is incomplete, both the amplitudes and the phases are corrected. The reconstruction also depends on number of observed and desired harmonics. Note also that the method is *non*-iterative.

The variance of the residual from the optimal reconstruction is given by:

$$\langle |\hat{a}_{lm} - a_{lm}|^2 \rangle = \left\{\frac{\langle \sigma_a^2 \rangle}{\langle a_l^2 \rangle_{th} + \langle \sigma_a^2 \rangle}\right\} \langle a_l^2 \rangle_{th}. \qquad (8)$$

The scatter is independent of the estimated optimal reconstruction. In the limit of negligible shot-noise the scatter vanishes and the reconstruction is deterministic. However, if $\langle a_l^2 \rangle_{th} \ll \langle \sigma_a^2 \rangle$ then the statistical scatter is that predicted by the *a priori* cosmic power-spectrum $\langle a_l^2 \rangle_{th}$.

### 3.3. Singular Value Decomposition

We solve eq. (6) by writing it as $\mathbf{c_{obs}} = \mathbf{B}\hat{\mathbf{a}}$, where $\mathbf{B} \equiv \mathbf{WF}^{-1}$, and by applying a Singular Value Decomposition (SVD) algorithm (e.g. Press et al. 1992). Briefly, the matrix $\mathbf{B}$ (of arbitrary dimensions) can be decomposed as $\mathbf{B} = \mathbf{U}\ diag\{\lambda_i\}\ \mathbf{V}^T$, where both $\mathbf{U}$ and $\mathbf{V}$ are orthonormal, and $\lambda_i$'s are the singular values of $\mathbf{B}$. The least-square solution is $\hat{\mathbf{a}} = \mathbf{V}\ diag\{1/\lambda_i\}\ \mathbf{U}^T\ \mathbf{c_{obs}}$. The singular values $\lambda_i$ give insight into the amount of useful information in the problem. To ensure stability of the reconstruction it is essential to set to zero $\lambda_i$'s which are much smaller than the maximal $\lambda_{max}$, before inverting by 'backsubstitution' (see Press et al. 1992). We find that in our problem of a $|b| = 5^o$ mask (see below) the inversion is rather insensitive to this truncation level (which controls the amount of regularization), but it is of great importance in the case of a larger mask (suggesting the need for an extra regularization). Other recent applications of the SVD approach in Astronomy include the analysis of galaxy spectra (Rix & White 1992) and helioseismology (Christensen-Dalsgaard et al. 1993).

### 4. APPLICATION TO IRAS 1.2 Jy DATA

Here we apply the method to the sample of IRAS galaxies brighter than 1.2 Jy (Strauss et al. 1992, Fisher 1992) which includes 5313 galaxies, and covers 88 % of the sky. This incomplete sky coverage is mainly due to the Zone of Avoidance, which we model as a 'sharp mask' at Galactic latitude $|b| < 5^o$. The mean number of galaxies is $\mathcal{N} = 392$ per steradian, which sets the shot-noise, $\langle \sigma_a^2 \rangle$. As our model for the cosmic scatter $\langle a_l^2 \rangle_{th}$ we adopt a fit to the observed power spectrum of IRAS galaxies (Fisher et al. 1993a) which is described (empirically) by a Cold Dark Matter model with a shape parameter $\Gamma = 0.2$, and with real-space normalization specified by the rms fluctuation in the number of IRAS galaxies in 8 $h^{-1}$ Mpc spheres, $\sigma_8 = 0.7$ . The Wiener filter (eq. 7) in this case drops monotonically with $l$, e.g. $\left\{\frac{\langle a_l^2\rangle_{th}}{\langle a_l^2\rangle_{th}+\langle\sigma_a^2\rangle}\right\} \sim 0.9, 0.7, 0.6$ and $0.3$ for $l = 1, 10, 15$ and $30$ respectively.

Figure 1 shows the reconstruction of the raw 2-D IRAS 1.2 Jy sample in Aitoff projection (in Galactic coordinates) for harmonics $1 \leq l \leq 15$. The Zone of Avoidance was left empty, and clearly it 'breaks' the possible chain of the Supergalactic Plane and other structures.

Figure 2 shows our optimal reconstruction where we have used observed harmonics $c_{lm}$'s with $1 \leq l \leq 15$ (255 independent coefficients in total) and reconstructed the whole-sky $a_{lm}$'s also for $1 \leq l \leq 15$. Now the structure is seen to be connected across the Zone of Avoidance, in particular in the regions of Centaurus/Great Attractor ($l \sim 315^o$), Hydra ($l \sim 275^o$) and Perseus-Pisces ($l \sim 315^o$), confirming the connectivity of the Supergalactic Plane. We also see the Puppis cluster ($l \sim 240^o$) recovered behind the Galactic Plane. This cluster has been noticed in earlier harmonic expansion (Scharf et al. 1992) and further studies (Lahav et al. 1993 and references therein). The other important feature of our reconstruction is the removal of shot noise all over the sky. This is particularly important for judging the reality of clusters and voids.

We have also compared our reconstruction with the one applied (using a $4\pi$ Wiener filter) to the IRAS sample in which the Zone of Avoidance was filled-in by extrapolating 'by hand' across the Galactic Plane (Yahil et al. 1991). The reconstructions look very similar both visually, and by $\chi^2$ and cross-correlation measures. We also found good agreement in the angular power-spectrum of the two reconstructions. As another test, we have used as a prior model the standard Cold Dark matter model (with $\Gamma = 0.5$), and found that the reconstructions changed very little.

As a more challenging test of the method we have also used an $N$-body simulation of standard Cold Dark matter (so the whole 'sky' true harmonics are known) and varied the size of the Zone of Avoidance. We find that for mask larger than $|b| = 15^o$ it is difficult to recover the unobserved structure. Clearly the success of the method depends on the interplay of three angular scales: the width of the mask, the desired resolution ($\pi/l_{max}$) and the physical correlation of structure.

## 5 DISCUSSION

In this *Letter* we have presented a Wiener filter method of reconstructing the full sky galaxy distribution and removing the shot noise. We have also shown that a variety of statistical approaches to the problem all lead to the same optimal Wiener estimator. The prior assumptions only depend on the observed 2-point galaxy correlation function and the nature of the shot-noise. While well known in image and signal processing, the method has not been implemented before (to our knowledge) in reconstructing the large scale structure in both amplitude and phase.

The 2-D spherical harmonics presentation and Wiener filtering approach can be also applied to other cosmic phenomena, e.g. to maps of the COBE microwave background and the HEAO1 X-ray background. Klypin et al. (1992) have used a similar regularization approach to analyze their *Relikt* map of the Microwave Background. However, their Tikhonov regularization procedure does not reflect the physical correlation of the underlying fluctuations.

Currently, we are developing the method further to 3-D by expanding the density field in spherical harmonics and spherical Bessel functions $j_l(kr)$ :

$$\rho(\mathbf{r}) - \bar{\rho} = \sum_{l>0} \sum_{m} \sum_{n} \rho_{lmn} \; j_l(k_{nl}\, r) \; Y_{lm}(\hat{\mathbf{r}}) \,, \qquad (9)$$

(e.g. Binney & Quinn 1991, and Lahav 1993 for a preliminary application to the 2 Jy redshift survey). This expansion generalizes the Wiener filter to handle the radial selection function and redshift distortion in harmonics (see Fisher et al. 1993b), as well as the shot-noise and incomplete sky coverage. The Wiener method is not limited to spherical harmonics or orthonormal set of functions, and can be implemented in Cartesian presentation as well (e.g. Hoffman 1993). Our procedure will be applied to new all-sky IRAS and optical redshift surveys, and to surveys of the peculiar velocity field. The 3-D noise-free $\rho_{lmn}$ coefficients will allow objective (non-parametric) comparison of different surveys of light and mass in the local universe.

**Acknowledgements**. We thank A. Dekel, D. Lynden-Bell, S. White and A. Yahil for stimulating discussions and M. Davis, J. Huchra, M. Strauss and A. Yahil for providing us with the IRAS data and simulations. OL thanks the Center for Micro and Macro Physics of the Hebrew University for the hospitality. SZ and YH acknowledge the hospitality of the Institute of Astronomy. CAS and KBF acknowledge SERC studentship and fellowship respectively.

**FIGURE CAPTIONS**

**Figure 1.** Harmonic expansion ($1 \leq l \leq 15$) of the projected raw IRAS 1.2Jy data in Galactic Aitoff projection. Regions not observed, in particular $|b| < 5$ (marked by dashed lines), were left empty. The contour levels of the projected surface number density are in steps of 100 galaxies per steradian (the mean projected density is $\mathcal{N} \sim 400$ galaxies per steradian).

**Figure 2.** A $4\pi$ Wiener reconstruction of the 2-D 1.2 Jy IRAS galaxy sample, for Harmonics $1 \leq l \leq 15$, plotted in Aitoff Galactic projection. The reconstruction corrects for incomplete sky coverage, as well as for the shot-noise. The assumed prior model is an empirical fit to the observed power-spectrum of IRAS galaxies. The reconstruction indicates that the Supergalactic Plane is connected across the Galactic Plane at Galactic longitude $l \sim 135^o$ and $l \sim 315^o$. The Puppis cluster stands out at the Galactic Plane at $l \sim 240^o$. The horizontal dashed lines at $b = \pm 5^o$ mark the major Zone of Avoidance in the IRAS sample. the contour levels of the projected surface number density are in steps of 100 galaxies per steradian (the mean projected density is $\mathcal{N} \sim 400$ galaxies per steradian).

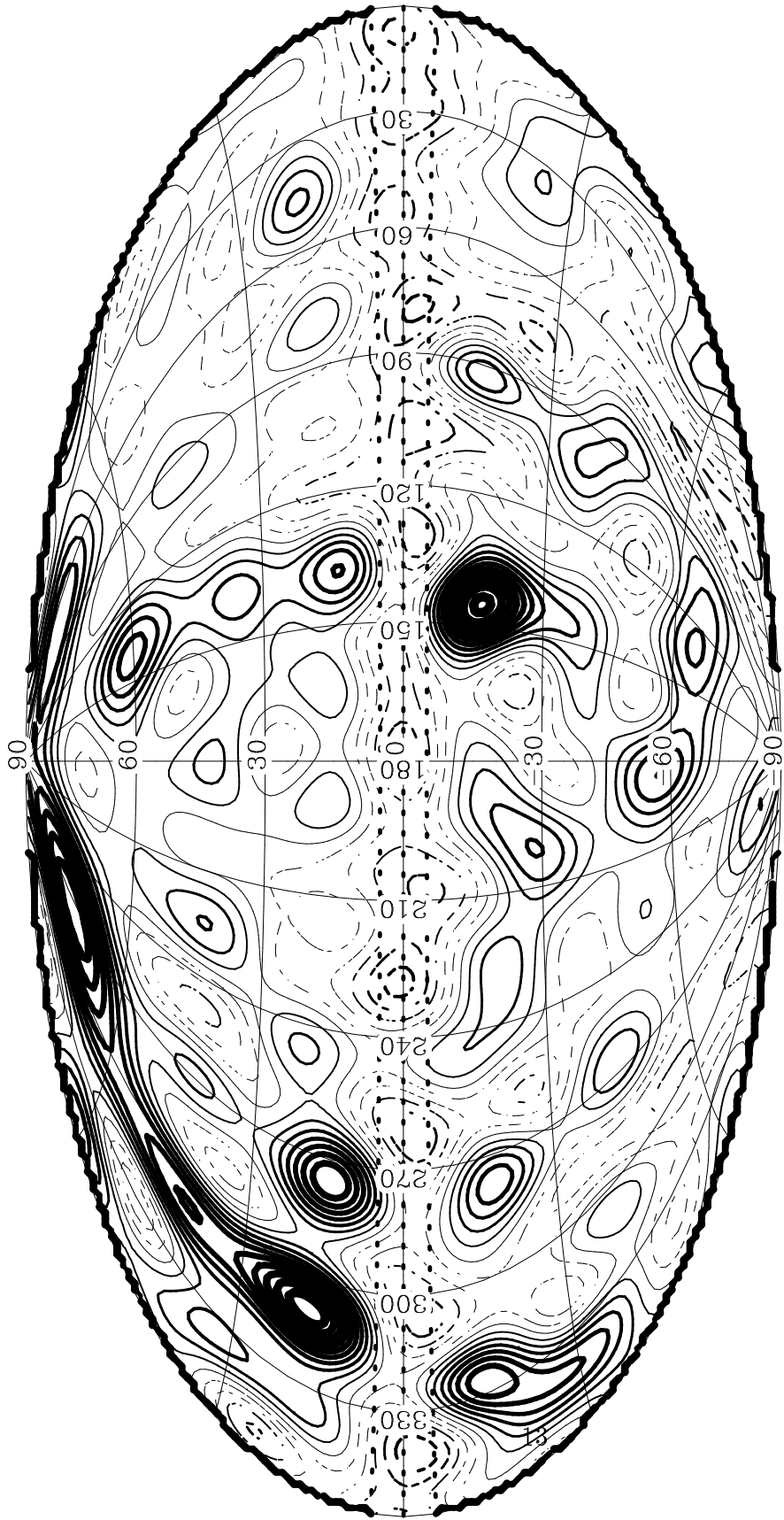

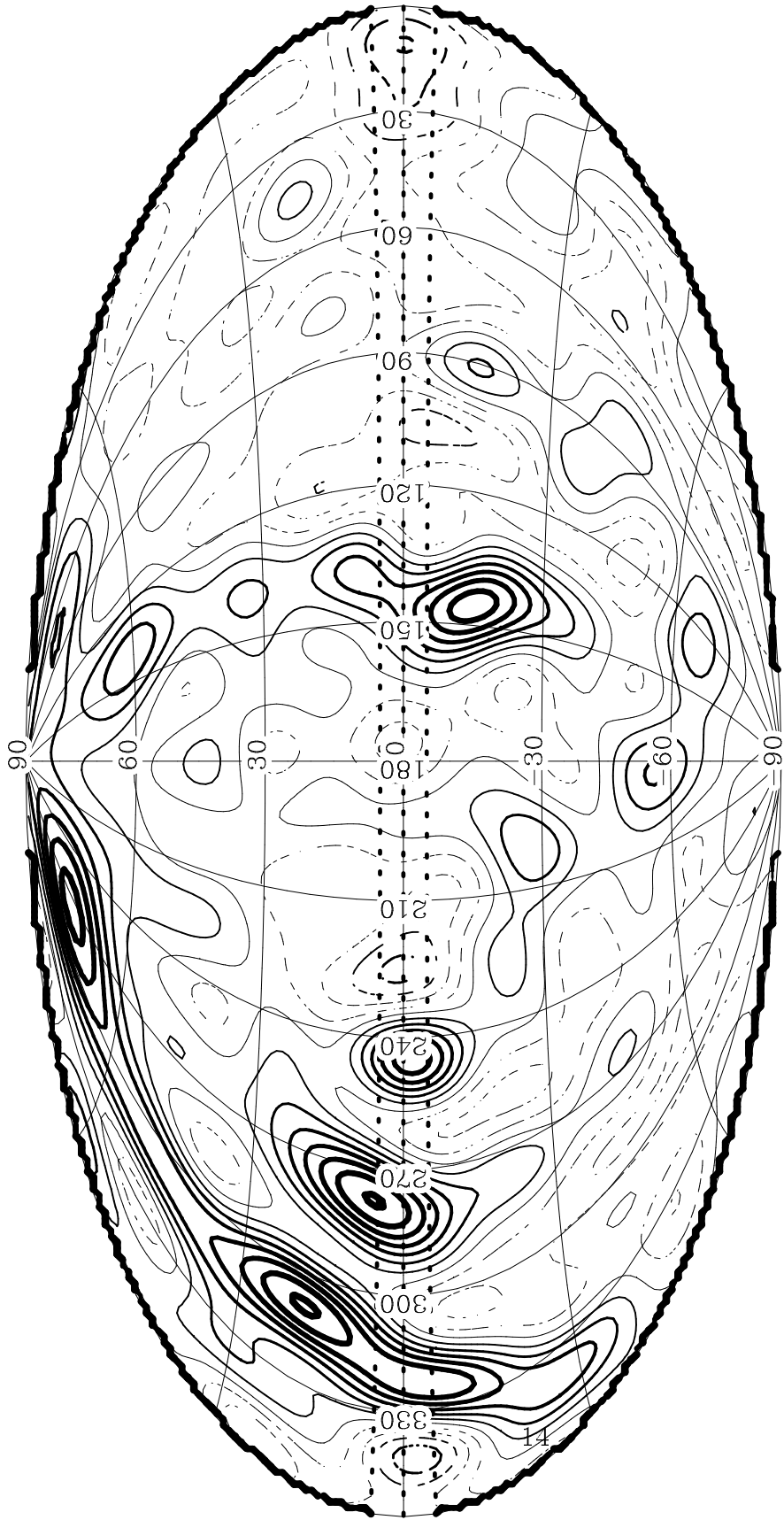

14